\documentclass[useAMS,usenatbib]{mn2e}
\usepackage{graphics,rotating,multirow,bm,color,hyperref,amsmath}
\hypersetup{
    colorlinks=true,
    linkcolor=black,
    citecolor=black,
}
\usepackage{dsfont}
\usepackage{amssymb}
\usepackage{graphicx}
\usepackage{verbatim}
\usepackage{natbib}
\usepackage{eucal}
\usepackage{calligra}

\usepackage{amsfonts}
\usepackage{color}
\usepackage[normalem]{ulem}
\usepackage[T1]{fontenc}
\DeclareMathAlphabet{\mathscr}{OT1}{pzc}%
                                 {m}{it}

\usepackage{times,epstopdf}
\usepackage{epsfig}
\usepackage[usenames,dvipsnames]{xcolor}
\usepackage{url}

\title[Eulerian Voids in Modified Gravity]
{Voids in Modified Gravity Reloaded: Eulerian Void Assignment}
\author[Lam et.~al.]
  {Tsz Yan Lam$^{1}$\thanks{tszylam@mpa-garching.mpg.de}, Joseph Clampitt$^2$\thanks{clampitt@sas.upenn.edu}, Yan-Chuan Cai$^3$\thanks{y.c.cai@durham.ac.uk}, Baojiu~Li$^{3}$\thanks{baojiu.li@durham.ac.uk}\\
  $^1$Max Planck Institute for Astrophysics, Karl-Schwarzschild-Str. 1, 85748 Garching, Germany\\
  $^2$Center for Particle Cosmology, Department of Physics and Astronomy, University of Pennsylvania, 209 S. 33rd St., Philadelphia, PA 19104, USA\\
  $^3$Institute for Computational Cosmology, Department of Physics, Durham University, South Road, Durham DH1 3LE, UK}
  
\pagerange{\pageref{firstpage}--\pageref{lastpage}} \pubyear{2014}

\def\LaTeX{L\kern-.36em\raise.3ex\hbox{a}\kern-.15em
    T\kern-.1667em\lower.7ex\hbox{E}\kern-.125emX}

\newcommand{\mnras}{MNRAS}

\newcommand{\apj}{ApJ}
\newcommand{\apjl}{ApJL}

\newcommand{\be}{\begin{eqnarray}}
\newcommand{\ee}{\end{eqnarray}}
\newcommand{\bes}{\begin{equation*}}
\newcommand{\ees}{\end{equation*}}
\newcommand{\bea}{\begin{eqnarray}}
\newcommand{\eea}{\end{eqnarray}}
\newcommand{\beas}{\begin{eqnarray*}}
\newcommand{\eeas}{\end{eqnarray*}}

\newcommand{\de}{{\rm d}}
\newcommand{\mpch}{{\rm Mpc}/h}

\newcommand{\rhoin}{\rho_{\rm in}}
\newcommand{\rhoout}{\rho_{\rm out}}

\newcommand{\phiout}{\phi_{\rm out}}

\newcommand{\lamin}{\lambda_{\rm in}}
\newcommand{\lamout}{\lambda_{\rm out}}

\newcommand{\mpl}{M_{\rm Pl}}

\newcommand{\reul}{R_{\rm E}}

\begin{document}

\label{firstpage}

\maketitle

\begin{abstract}
{We revisit the excursion set approach to calculate void abundances in chameleon-type modified gravity theories, which was previously studied by \citet{ccl2013}. We focus on properly accounting for the void-in-cloud effect, i.e., the growth of those voids sitting in over-dense regions may be restricted by the evolution of their surroundings. This effect may change the distribution function of voids hence affect predictions on the differences between modified gravity and GR. We show that the thin-shell approximation usually used to calculate the fifth force is qualitatively good but quantitatively inaccurate. Therefore, it is necessary to numerically solve the fifth force in both over-dense and under-dense regions. We then generalise the Eulerian void assignment method of \citet{pls2012b} to our modified gravity model.  We implement this method in our Monte Carlo simulations and compare its results with the original Lagrangian methods. We find that the abundances of small voids are significantly reduced in both modified gravity and GR due to the restriction of environments. However, the change in void abundances for the range of void radii of interest for both models is similar.  Therefore, the difference between models remains similar to the results from the Lagrangian method, especially if correlated steps of the random walks are used. As \citet{ccl2013}, we find that the void abundance is much more sensitive to modified gravity than halo abundances. Our method can then be a faster alternative to N-body simulations for studying the qualitative behaviour of a broad class of theories. We also discuss the limitations and other practical issues associated with its applications.}
\end{abstract}


\section{Introduction}

\label{sect:intro}
The discovery of the accelerated expansion of the Universe 
sparked a surge of research on the possibility of modified
gravity models \citep[see, for example,][for reviews]{jk10,cliftonetal}.
The main goal of such modifications is to alter the large
scale behaviour to explain the acceleration -- 
however, any modifications in the gravity model must at the same time 
satisfy the tight constraints from the solar system tests.
One way to fulfil this requirement is to include some 
kind of screening mechanism to suppress the modification in the
local environment (i.e., regimes with high matter density or deep Newtonian potential). In this work we focus on one particular subclass of modified gravity theories -- one that modifies
gravity by introducing a dynamical scalar field that mediates a fifth force. 

The chameleon model of \citet{kw} is a very representative example of this class of modified gravity theories. {In this model, the acceleration of the Universe is associated with a scalar field, which has a runaway type self-interaction potential and an interaction (coupling) with matter. Such a specific setup ensures that during the cosmic evolution the scalar field is trapped to the vicinity of $0$, so that its potential energy roughly stays as a constant -- of order the energy density of the cosmological constant in the $\Lambda$CDM paradigm -- which means that the expansion history can be very close to that of $\Lambda$CDM. The coupling to matter produces a fifth force which can be as strong as standard gravity, so that the model is naively ruled out by solar system tests of gravity. However, because the scalar field is trapped close to 0 (especially in regions with high matter density), the fifth force can be severely suppressed, thereby passing the stringent local constraints.} {This is the chameleon mechanism \citep{kw}, in which the behaviour of the fifth force depends on the environment: the suppression of the fifth force in the solar system is achieved by reducing its force range to sub-millimetres, while on the cosmological background (at late times) the force range can be of order $\mathcal{O}(1-10)$~Mpc, on which scales strong deviations from $\Lambda$CDM can be found. Notice that on even larger scales, the effect of the fifth force diminishes, implying that the effect on very large-scale structures is minimal.} 

{Because the} background evolution and the linear perturbations on large scales 
can be indistinguishable from the 
standard $\Lambda$CDM cosmology \citep{hs2007,lb2007,lz2009}{, and}
solar system tests are satisfied by construction{, nonlinear} structure formation {(on scales of $10^0\sim10^2$~Mpc)} is the only regime where
effects of such models would possibly be detected.
A number of studies 
\citep[e.g.,][]{oyaizu2008,olh2008,sloh2009,lz2009,lz2010,lb2011,zlk2011,lztk2011}
employed N-body numerical simulations to study nonlinear structure formation
-- however high resolution simulations with cosmological volume 
are still challenging due to the highly non-linear equation governing the 
scalar field. {As a result, semi-analytic methods are often used to make qualitative predictions and provide insights into the underlying physics for such models.}

This work {will be along this direction and} aim at investigating the effect of the chameleon-type modified gravity on 
the large-scale structure {in the nonlinear regime}. We will extend the standard excursion set approach \citep{bcek,mw1996,st1999} to predict the abundance of structures.
The application of this approach to halo abundance and bias 
have been visited by \citet{le2012,lilam2012,lamli2012}, and therefore in this work we will focus on cosmic voids.

Cosmic voids found in redshift surveys have many applications. Observationally, \citet{hvp2012} studied the photometric properties of void galaxies found by \citet{hv2002,pvh2012}. 
There is also ongoing work toward using voids as cosmological probes. The Alcock-Paczynski test has been proposed \citep{lw2012} and progress has even 
been made towards applying it to data \citep{slw2012b, spw2014}. Stacking of voids for the cosmic microwave background has been used to detect the integrated Sachs Wolfe effect as an alternative 
to the cross-correlation method \citep{Granett2008, Cai2014a, Cai2014b, Hotchkiss2014, Ilic2013, Planck2013}; 
Void-void and void-galaxy clustering is another promising cosmological probe \citep{hws2014}. Recent measurements of the weak lensing signal \citep{mss2013} 
and density profile of voids \citep{cj2014} gives a direct handle on their dark matter content. Void ellipticity has been shown to be sensitive to the dark energy 
equation of state \citep{Bos2012}. Voids properties have been studied in coupled dark energy model using N-body simulations \citep{li2011, Sutter2014}.

{The excursion set predictions of void abundances have been done previously}
by \citet{ccl2013}, where they showed that voids, due to the under-density {and therefore weaker suppression of the fifth force inside them}, 
would yield a stronger signal of modified gravity compared to halos. 
This work focuses on applying a modern void-assignment algorithm, Eulerian-void assignment,
in the context of the same modified gravity model. Eulerian-void assignment was recently 
proposed by \citet{pls2012b} as an improvement to the traditional Lagrangian-based 
assignment by taking into account the effect of the immediate surroundings on the growth of voids. 

This paper is organised as follows: In \S~\ref{sect:chameleon} 
we briefly review the theoretical model to be considered and summarise
its main ingredients.  {In \S~\ref{sect:force_solutions} we find and validate the numerical solutions to the fifth force, which will be used in \S~\ref{sec:evolution} to calculate numerically the evolution of spherical over- and under-densities in the modified gravity model. \S~\ref{sect:method} presents a detailed description of the different void assignment methods adopted in the literature and in this paper; in particular, we will describe how the Eulerian method of \citet{pls2012b} can be implemented in the context of modified gravity. In \S~\ref{sect:results} we show the comparison of predictions on the void abundance from the different methods and demonstrate their consistency. Finally, \S~\ref{sect:summary} summarises the main results and discusses their implications.}

%
\section{The Chameleon Theory}
\label{sect:chameleon}

This section presents the theoretical framework for investigating the cosmological effects of a coupled scalar field.  We will give the relevant general field equations in \S~\ref{subsect:equations}, and then
specify the models analysed in this paper in \S~\ref{subsect:specification}.

\subsection{Cosmology with a coupled scalar field}
\label{subsect:equations}

The equations governing the scalar field can be found in \citet{lz2009, lz2010, lb2011}, and  are presented here only to make this paper self-contained. Because of this, this section is brief and contains only the most essential ingredients of coupled scalar field cosmology. Interested readers are referred to the above references for more details.

We start from a Lagrangian density
\begin{eqnarray}\label{eq:lagrangian}
{\cal L} =
{1\over{2}}\left[\mpl^{2}{R}-\nabla^{a}\phi\nabla_{a}\phi\right]
+V(\phi) - C(\phi) ({\cal L}_{\rm{DM}} + {\cal L}_{\rm{S}}),
\end{eqnarray}
in which $R$ is the Ricci scalar, the reduced Planck mass is $\mpl=1/\sqrt{8\pi G}$ with $G$ being Newton's constant, and ${\cal L}_{\rm{DM}}$ and
${\cal L}_{\rm{S}}$ are respectively the Lagrangian
densities for dark matter and standard model fields. $\phi$ is
the scalar field and $V(\phi)$ is the potential describing its self interaction. The coupling
function $C(\phi)$, on the other hand, describes its interaction with matter. The coupled scalar field model is then fully specified by the functional forms for $V(\phi)$ and $C(\phi)$.

Varying the total action with respect to the metric $g_{ab}$, we
obtain the following expression for the total energy momentum
tensor in this model:
\begin{eqnarray}\label{eq:emt}
T_{ab} = \nabla_a\phi\nabla_b\phi -
g_{ab}\left[{1\over2}\nabla^{c}\nabla_{c}\phi-V(\phi)\right]+ C(\phi) (T^{\rm{DM}}_{ab} + T^{\rm{S}}_{ab}),\nonumber
\end{eqnarray}
where $T^{\rm{DM}}_{ab}$ and $T^{\rm{S}}_{ab}$ are the
energy momentum tensors for (uncoupled) dark matter and standard
model fields. The existence of the scalar field and its coupling
change the form of the energy momentum tensor, leading to
potential changes in the background cosmology and 
structure formation.

The coupling to the scalar field generates an extra
interaction (the fifth force) between matter
particles, which can be regarded as a result of the exchange of scalar quanta. This is best
illustrated by the geodesic equation for dark matter particles
\begin{eqnarray}\label{eq:geodesic}
{{{\rm d}^{2}\bf{r}}\over{{\rm d}t^2}} = -\vec{\nabla}\Phi -
{{C_\phi(\phi)}\over{C(\phi)}}\vec{\nabla}\phi,
\end{eqnarray}
where $\bf{r}$ is the particle position, $t$ the physical time, $\Phi$
the Newtonian potential and $\vec{\nabla}$ is the spatial
derivative; $C_\phi\equiv {\rm d}C/{\rm d}\phi$. The second term on the
right hand side is the fifth force, whose potential (the conservative force potential, not to be confused with the self-interacting potential $V(\phi)$) can be described by $\ln C(\phi)$.

Equation (\ref{eq:geodesic}) suggest that to follow the motion of particles we need to know the time
evolution and spatial configuration of $\phi$. This is usually achieved by explicitly solving the scalar field equation of motion
\begin{eqnarray}
\nabla^{a}\nabla_a\phi + {{\rm{d}V(\phi)}\over{\rm{d}\phi}} +
\rho {\frac{\rm{d}C(\phi)}{\rm{d}\phi}} = 0,
\end{eqnarray}
where $\rho = \rho_{\rm DM} + \rho_{b}$ is the sum of dark matter and
baryonic matter densities. Equivalently, the equation of motion can be witten as
\begin{eqnarray} \label{eq:eom0}
\nabla^{a}\nabla_a\phi + {{\rm{d}V_{\rm eff}(\phi)}\over{\rm{d}\phi}} = 0,
\end{eqnarray}
where we have defined an effective potential for the scalar field
\begin{eqnarray}\label{eq:Veff}
V_{\rm eff}(\phi) = V(\phi) + \rho C(\phi).
\end{eqnarray}
The background evolution of $\phi$ can be solved given
the present-day value of  $\bar{\rho}$ (note that 
$\bar{\rho}\propto a^{-3}$ \citep{lz2009}). We can then split $\phi$
into two parts, $\phi=\bar{\phi}+\delta\phi$, where
$\bar{\phi}$ is the background value of $\phi$ and $\delta\phi$ is its
perturbation around $\bar{\phi}$, and subtract the
background part of the scalar field equation of motion from the full equation
to find the equation of motion for $\delta\phi$. In the
quasi-static limit, where we can neglect all time derivatives of
$\delta\phi$ compared with its spatial derivatives (this
is a good approximation on scales well within the horizon),
this can be obtained as
\begin{eqnarray}\label{eq:scalar_eom}
\vec{\nabla}^{2} \delta \phi =
{{\rm{d}C(\phi)}\over{\rm{d}\phi}}\rho -
{{\rm{d}C(\bar{\phi})}\over{\rm{d}\bar{\phi}}}\bar{\rho} +
{{\rm{d}V(\phi)}\over{\rm{d}\phi}} -
{{\rm{d}V(\bar{\phi})}\over{\rm{d}\bar{\phi}}}.
\end{eqnarray}

The calculation of the scalar field $\phi$ using the above equation then completes the source term of the Poisson equation,
\begin{eqnarray}\label{eq:poisson}
\vec{\nabla}^{2} \Phi &=& \frac{1}{2 \mpl^2}\left[\rho_{\rm tot}+3p_{\rm tot}\right] - \frac{1}{2 \mpl^2}\left[\bar{\rho}_{\rm tot}+3\bar{p}_{\rm tot}\right]\nonumber\\
&=& {{1}\over{2 \mpl^2}}\left[\rho C(\phi) - \bar{\rho}C(\bar{\phi}) - 2V(\phi) + 2V(\bar{\phi})\right], 
\end{eqnarray}
where we have neglected the kinetic energy of the scalar field since the running of $\phi$ is always negligible for the model studied here. In Eq.~(\ref{eq:poisson}), we have again exploited the quasi-static approximation by dropping time derivatives of the Newtonian potential.

\subsection{Specification of Model}
\label{subsect:specification}

As mentioned above, to completely specify the coupled scalar field model, we need to give the functional forms of $V(\phi)$ and $C(\phi)$. Here, for illustration purpose, we will use the chameleon-type models investigated by \cite{lz2009, lz2010}, with
\begin{eqnarray}\label{eq:coupling}
C(\phi) = \exp(\gamma \phi / \mpl),
\end{eqnarray}
and 
\begin{eqnarray}\label{eq:pot_chameleon}
V(\phi) = {{\rho_{\Lambda}}\over{\left[1-\exp\left(-\phi / \mpl \right)\right]^\alpha}}.
\end{eqnarray}
In the above $\rho_{\Lambda}$ is the energy density of the cosmological constant in the standard $\Lambda$CDM scenario, $\phi$ plays the role of dark energy in this model, and $\gamma, \alpha$ are dimensionless  model parameters which control the strength of the 
coupling $C(\phi)$ and the slope of the scalar field self-interaction potential $V(\phi)$ respectively. 

We choose $\alpha\ll1$ and $\gamma>0$ as in \citet{lz2009, lz2010},
so that the global minimum of $V_{\rm eff}(\phi)$ is always very close to $\phi=0$ throughout the cosmic evolution,
and that $m^2_{\phi} \equiv {\rm d}^2V_{\rm eff}(\phi)/{\rm d}\phi^2$ at this
minimum is very large in high-density regimes. For example, the minimum of $V_{\rm eff}(\phi)$ on a background with matter density $\rho$ is approximately \citep{lz2009}
\begin{eqnarray}
\phi = \frac{\alpha\rho_{\Lambda}}{\gamma\rho},
\end{eqnarray}
such that $\phi\ll1$, especially in high density regions where $\rho\gg\rho_{\Lambda}$. Such choices of model parameters ensure that: 

\indent$\bullet$ $\phi$ is
trapped in the vicinity of $\phi=0$ throughout the cosmic history and thus
$V(\phi)\approx \rho_{\Lambda}$ behaves as a cosmological constant. For this reason,
we take the background expansion of this model to be exactly the same as that of $\Lambda$CDM with the same cosmological parameters. Note that this is not guaranteed if $\alpha\ll1$ does not hold.

\indent$\bullet$ the fifth force is strongly
suppressed in high density regions where $\phi$ acquires a large
mass, $m^2_{\phi}\gg H^2$ ($H$ is the Hubble expansion rate),
and therefore the fifth force cannot propagate a long distance without decaying. This is because the fifth force,
mediated by a scalar field, takes the Yukawa form and decays exponentially
over the Compton wavelength, $\lambda\equiv m^{-1}_{\rm \phi}$, of this scalar field.

The fact that the fifth force is strongly suppressed when matter density is high implies that its influence on
structure formation occurs mainly at late times.
The environment-dependent behaviour of the fifth force was
first considered in \citet{kw}, and has since then been known
as `chameleon screening'. It is one of the most well-studied modified gravity theories: because of the finite range of 
the fifth force, and because of the severe suppression of it in high-density regions,
it is believed that the strongest cosmological constraints on such models come from the 
study of cosmic voids, which are low density regions ($\delta\sim-0.8$) in the Universe with sizes ranging from a few to $\mathcal{O}(100)$ Mpc.

\section{Force solutions}
\label{sect:force_solutions}

Although in this work we are only concerned with void abundances, the Eulerian void assignment method of \citet{pls2012b} requires solutions for collapsing over-dense walls around the voids. Thus, the equations below will apply generally to calculate fifth forces for under- {\it and} over-densities. The distinction between the two regimes is made entirely in the choice of parameters, specifically the ratio of the object's density to that of its environment. After presenting the general equations, we will focus on the over-dense solutions since they are unique to this work (under-dense solutions were already shown in detail in \citet{ccl2013}). Finally, we will describe numerical checks of our algorithm.

\subsection{Scalar field solution}
\label{sec:scalar}

We are interested in the simplest model of a dark matter halo {(or void)}, with top-hat {(or bucket)} density profiles described by
\be \label{eq:density}
\rho_0 (\chi) = \left\{
\begin{matrix}
\rhoin \qquad {\rm ~~for} \qquad \chi \le r \cr
\rhoout \qquad {\rm for} \qquad \chi > r
\end{matrix}
\right. \, ,
\ee
in which $r$ is the object's radius {and $\chi$ is a variable characterising the distance from the centre of the object. Because $\phi$ is confined to the vicinity of $\phi=0$, we have} $C(\phi) \approx 1$ and the first term on the right-hand side of Eq.~(\ref{eq:poisson})
can be integrated once to find the force per unit test mass
\be
F_{\rm N} (\chi) = -\frac{G M(< \chi)}{\chi^2} \, .
\ee
The second term on the right-hand side of Eq.~(\ref{eq:geodesic}) is the fifth force, which for our choice of $C(\phi)$ in Eq.~(\ref{eq:coupling}) can be expressed as
\be \label{eq:F_5}
F_\phi (\chi) = -\gamma \frac{\de}{\de \chi} (\phi / \mpl) \, .
\ee
We define the ratio of fifth to Newtonian forces {at the surface of the object ($\chi=r$)} as
\bea
\eta\ \equiv\ \frac{F_\phi}{F_{\rm N}} \ =\ \frac{6 \gamma \mpl}{r \rhoin} \left. \frac{\de \phi}{\de \chi}\right|_{\chi = r} \, , \label{eq:eta}
\eea
which is constrained to be $\eta \le 2\gamma^2$ for over-densities.

{In the model considered in this work, the force ratio $\eta$ can be equivalently determined by three length scales: the radius of the object $r$, and the Compton wavelengths inside and outside the object. The latter are given by the following relations:}
\be
\rho_i = \left(\frac{\mpl \sqrt{\alpha \rho_{\Lambda}}}{\gamma}\right) \frac{1}{\lambda_i} \,
\ee
where $i = $ [in, out] and $\lambda$ is the Compton wavelength of the scalar field (note that the Compton wavelength is small on a high matter-density background and vice versa). In \citet{ccl2013}, we showed that the degrees of freedom of a spherical top-hat under-density can be reduced from the three length scales, $r$, $\lamin$, and $\lamout$, to two ratios of these lengths: $r/\lamout$ and $\lamout/\lamin$. This logic proceeds without change for over-densities, and so we only quote the key equations here.

The effective potential for the specific model studied here can be written as
\be \label{eq:eff}
V_{\rm eff} (\phi) = {{\Lambda}\over{\left[1-\exp\left(-\phi / \mpl \right)\right]^\alpha}} + \rho_i \exp(\gamma \phi / \mpl) \, ,
\ee
with which the equation of motion Eq.~(\ref{eq:eom0}) can be simplified to read
\be \label{eq:eom}
\frac{\de^2 \psi}{\de \tau^2} + \frac{2}{\tau}\frac{\de \psi}{\de \tau} + \frac{1}{\psi} = \left\{
\begin{matrix}
\lamout / \lamin \qquad {\rm for} \qquad \tau \le r / \lamout \cr
1 \qquad {\rm ~~~~~~~~~~~~~for} \qquad \tau > r / \lamout
\end{matrix}
\right.
\ee
with the following two boundary conditions
\be \label{eq:bc}
\left.\frac{\de \psi}{\de \tau}\right|_{\tau = 0} = 0, \qquad \qquad \psi(\tau \rightarrow \infty) = 1 \, .
\ee
Note that in the above the solution has been rescaled and written in terms of $\psi \equiv \phi / \phiout$ and $\tau \equiv \chi / \lamout$.
Rewriting Eq.~(\ref{eq:eta}) in terms of the new variables and using $\rho_\Lambda = \Omega_{\Lambda} \rho_{\rm c}$ {(where $\rho_{\rm c}$ is the critical density at the present day)}, we find
\be \label{eq:eta2}
\eta (r) = 6 \gamma \sqrt{\frac{\alpha \Omega_{\Lambda}}{\Omega_m}} \frac{\mpl \sqrt{\bar{\rho}_m}}{r \rhoin}  \left. \frac{\de \psi}{\de \tau}\right|_{\tau = r / \lamout} \, ,
\ee
where $\bar{\rho}_m$ is the background matter density today.

The recasting of the scalar field equation of motion in the form of Eq.~(\ref{eq:eom}) has the advantage that the fifth-force-to-gravity ratio $\eta(r)$ can then be obtained by interpolating a 2D pre-computed table, for spherical top-hat systems of arbitrary size, density, and in any kind of environment. This can in turn greatly simplify the numerical computations in this work, which would otherwise be prohibitively expensive.

\begin{figure}
\centering
\includegraphics[width=0.52\textwidth]{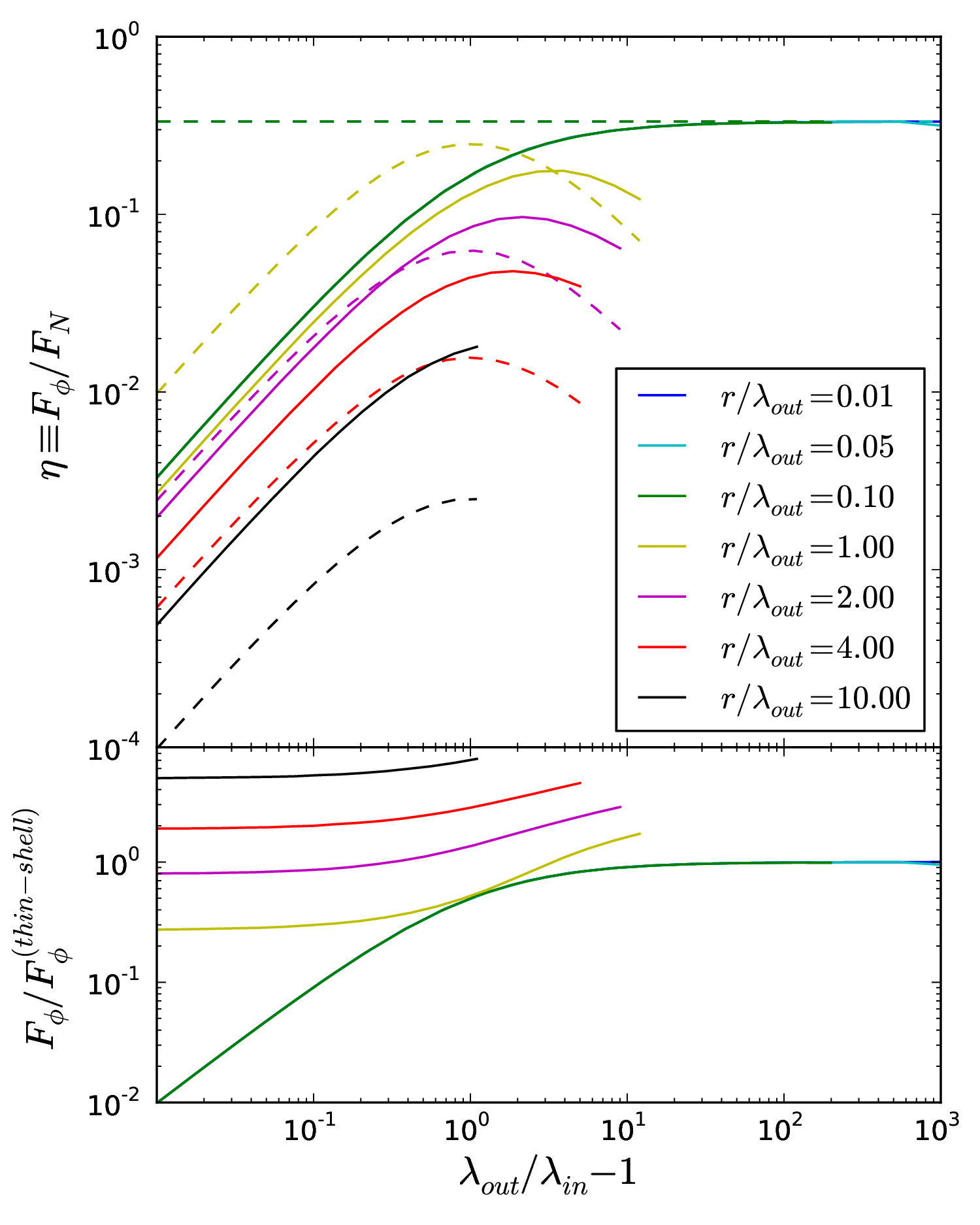}
\caption{({\it top panel}): We compare the value of the force ratio $\eta = F_\phi / F_N$ in both the exact solution (solid lines) and the thin-shell approximation (dotted lines). For $r / \lamout > 10$ the fifth force is less than 2\% of the strength of gravity. Thus we do not need the exact solution and can approximate the fifth force as zero there.
({\it bottom panel}): The ratio of the curves in the top panel. The approximate solution is accurate for small $r / \lamout$, large $\lamout / \lamin$ values, but otherwise does not reproduce well the exact result. At small $\lamout / \lamin$ however, the fifth force is quite weak anyway, so we can set it to zero safely.}
\label{fig:thin-shell}
\end{figure}

\subsection{Comparison to the thin-shell approximation}

The radial profile of a chameleon-type scalar field has been studied in detail for spherical over-densities {\citep[see, e.g.,][for a few examples]{kw,lkzl2012,dlss2012}}. \citet{kw} derived a simple analytical formula for the fifth force which has been shown to agree well with numerical simulations \citep{lzk2012}. Here we show that the thin-shell prescription followed by \citet{le2012} {(rewritten using our reduced parameters)} given by
\be
F_\phi^{\rm (thin-shell)} / F_{\rm N} = \frac{\lamin \lamout}{r^2} - \left(\frac{\lamin}{\lamout}\right)^2,
\ee
gives qualitatively {reasonable} results.

In Fig.~\ref{fig:thin-shell} we compare the two solutions, where the solid lines in the top panel show the numerical solutions for the ratio of forces $\eta$ as a function of the two dimensionless parameters {$r/\lambda_{\rm out}$ and $\lambda_{\rm out}/\lambda_{\rm in}$}, while the dotted lines show the thin-shell approximation. $\lambda_{\rm out}> \lambda_{in}$ means that the density inside the top-hat region is greater than outside, which is the case for haloes. Both solutions are qualitatively the same: 
\begin{itemize}
\item When the exterior and interior Compton wavelengths are very close in magnitude, the fifth force is negligible compared to gravity{, because the scalar field takes very similar values inside and outside the over-density, making its gradient (i.e, the fifth force) small.}
\item As $\lamout$ increases, the fifth force grows in importance until $\lamout \sim$ a few $\times \lamin$, after which {the ratio $\eta=F_\phi/F_{\rm N}$ decreases. This can be understood as follows: suppose $\lambda_{\rm in}$ remains unchanged, then an increase in $\lambda_{\rm out}$ with fixed $r/\lambda_{\rm out}$ means that the radius of the over-density, and so the Newtonian potential $\Phi_{\rm N}\propto r^2\propto\lambda^2_{\rm out}$, is increased; meanwhile, according to the thin-shell approximation, the fifth-force-to-gravity ratio is proportional to $|\phi_{\rm out}-\phi_{\rm in}|/\Phi_{\rm N}$, with $\phi_{\rm out}\approx\alpha\rho_{\Lambda}/\gamma\rho_{\rm out}\propto\lambda_{\rm out}$ for our model: when $\phi_{\rm out}$ is close to $\phi_{\rm in}$, i.e., when $\lambda_{\rm out}/\lambda_{\rm in}\sim1$, $|\phi_{\rm out}-\phi_{\rm in}|$ increases faster than $\Phi_{\rm N}$, making $\eta$ increase; but when with $\phi_{\rm out}\gg\phi_{\rm in}$, $\eta$ becomes $\propto$ $|\phi_{\rm out}/\Phi_{\rm N}|\propto\lambda^{-1}_{\rm out}$, which decreases with $\lambda_{\rm out}$.} Physically, there are two parts of the screening, self and environment screening: the self screening becomes stronger with increasing mass (and therefore increasing $r$), while the environment screening becomes weaker with decreasing environmental density (and therefore increasing $\lambda_{\rm out}$). The behaviour of $\eta$, with $r/\lamout$ fixed, is the result of the competition of the two.
\item The fifth force ratio {$\eta$ decreases} monotonically with {increasing radius $r$ of the spherical over-density}, for fixed $\lamout$. {This is because a larger $r$ means that the over-density is more massive, and therefore more efficient self-screening of the fifth force in given environment (specified by $\lamout$).}
\end{itemize}

While it is qualitatively very good, the thin-shell solution can be up to a factor of ten too large or too small, as shown by the ratio of exact to thin-shell solutions in the bottom panel of Fig.~\ref{fig:thin-shell}. The thin-shell solution is quantitatively correct only in the regime in which the exterior Compton wavelength is much larger than both the over-density radius and its interior Compton wavelength. This is where the fifth force has its largest magnitude relative to gravity, $F_\phi = F_{\rm N} / 3$, the thick-shell regime, {and for other regimes we need to go beyond the thin-shell approximation to be accurate}. Furthermore, because the Eulerian environment results of this work require the fifth force solution for both over- and under-densities {(and because the thin-shell approximation fails for the latter)}, we need the exact (numerical) solution in both cases for the sake of continuity and consistency.

\subsection{Validation of numerical solution}

\begin{figure}
\centering\includegraphics[width=0.52\textwidth]{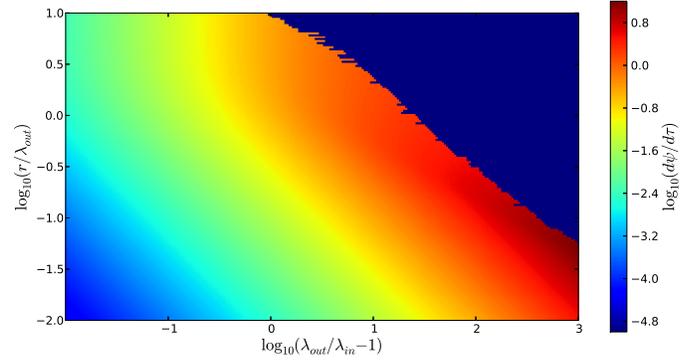}
\caption{The solution of Eq.~(\ref{eq:eom}) as a function of dimensionless parameters $r / \lamout$ and $\lamout / \lamin$, in the over-density case for which $\lamout / \lamin > 1$. The fifth force $F_\phi \propto \de \psi / \de \tau$ grows monotonically with either quantity, for all relevant parameter space. The top-right corner is numerically difficult to solve, but we have verified that this regime of parameter space is not used in obtaining the Eulerian barriers.}
\label{fig:force-table}
\end{figure}

Having demonstrated that a numerical solution of the fifth force is necessary for both under- and over-densities, we proceed to the non-trivial validation of such solutions. The void solutions of Eq.~(\ref{eq:eom}) correspond to the parameter space with $\lamout / \lamin < 1$. These are shown in \citet{ccl2013} and are numerically straightforward. However, for the over-dense case ($\lamout / \lamin > 1$), much of the parameter space is difficult to solve. In Fig.~\ref{fig:force-table} we show these solutions of Eq.~(\ref{eq:eom}) as a function of the two dimensionless parameters. In the top-right corner of this plot we were unable to obtain numerical solutions, and therefore set $\de \psi / \de \tau = 0$ for display purposes. However, according to Fig.~\ref{fig:thin-shell}, the fifth force can still be a significant fraction of gravity in this regime (more precisely, it can have a strength > 2\% of gravity.)

Fortuitously, over-densities with the range of Eulerian radii in which we are interested ($R_{\rm eul} \ge 5 \mpch$) do not stray into this unsolved regime.
The solutions of Sec.~\ref{sec:evolution} require calculating the evolution of an over-density from deep in the matter dominated era to the present day. 
Early in cosmic history the $r / \lamout$ parameter is very large (because both $r$ is large and $\lamout$ is quite small due to the high density) 
and $\lamout / \lamin$ is only slightly larger than one. Then, for collapsing over-dense regions $\lamout / \lamin$ grows monotonically with time, 
while $r / \lamout$ decreases monotonically with time. Qualitatively, then, all solutions track from the top left of the Fig.~\ref{fig:force-table} force table 
to the bottom right as time passes.

The difficulty of obtaining fifth force solutions is greater for larger objects, corresponding to larger smoothing radii in the initial Lagrangian density 
field or equivalently smaller $s$. (Here $s=\sigma^2$ is the variance of the smoothed linear matter power spectrum, to be defined below.)
However, we have verified that we can obtain the Eulerian barriers (again to be defined below) all the way to a small value of $s$, that is, down 
to some $s_{\rm min}$ such that
\begin{eqnarray}
\delta_{\rm lin} > 5 \sqrt{s_{\rm min}} = 5 \sigma \, .
\end{eqnarray}
We do not need the barrier solution at such small $s$ since they are seldom hit by random walks starting at the origin. In summary, we have verified that for each timestep with $F_\phi \ge 0.02 F_{\rm N}$, we solve the exact numerical solution. For timesteps with $F_\phi < 0.02 F_{\rm N}$, we set the fifth force to zero.

\section{Evolution of density perturbations}
\label{sec:evolution}

{In our simplified modelling of the evolution of spherical over- and under-densities, we make the assumption that the profile remains as a top-hat or bucket during the whole course and there are no shell crossings. Under this premise, w}e need to track the evolution of two shells: the halo {(for over-density cases)} or the void {(for under-density cases)} shell, with proper radius $r_{\rm in}(t)$, as well as {the shell of the environment that co-evolves with the over- or under-density inside it \citep{le2012}}, with radius $r_{\rm env}(t)$. The evolution equation is given by
\begin{eqnarray}\label{eq:revolution2}
\frac{\ddot{r}_j}{r_j} = -\frac{1}{6\mpl^2}\left[\rho_j (1+\eta)-2\rho_\Lambda\right].
\end{eqnarray}
where $\rho_j \equiv 3M_j/4\pi r_j^3$ is the matter density in the spherical shell and the constant $\rho_\Lambda\approx V(\phi)$ is the 
effective dark energy density defined in \S~\ref{sec:scalar}. The notation $r_j$ denotes $r_{\rm in}$ or $r_{\rm env}$.

The only difference between the evolution of the inner shell and that of its environment is the effect of the fifth force. For the inner shell we calculate the fifth force, whereas the larger environment shell is approximated as evolving under GR, $\eta = 0${, because the environment is generally taken to be larger than the Compton wavelength of the scalar field, for which the fifth-force-to-gravity ratio is small [cf.~Fig.~\ref{fig:thin-shell}]}. To calculate the fifth force on the inner shell at each time-step we use a spherical top-hat profile,
\be \label{eq:density-evo}
\rho (\chi) = \left\{
\begin{matrix}
\rho_{\rm in} \qquad {\rm ~~for} \qquad \chi \le r_{\rm in} \cr
\rho_{\rm env} \qquad {\rm for} \qquad \chi > r_{\rm in}
\end{matrix}
\right. \, .
\ee
The fifth-force-to-gravity ratio is then, using Eq.~(\ref{eq:eta2}),
\begin{eqnarray} \label{eq:eta-evo}
\eta = \frac{\sqrt{3\alpha\Omega_\Lambda}\gamma\frac{{\rm d}\psi}{{\rm d}\tau}\Big|_{\tau = r / \lambda_{\rm env}}}{\frac{1}{2}\Omega_m\left(H_0R\right)\left(ay_{\rm{in}}\right)^{-2}} \, ,
\end{eqnarray}
for the inner shell. {Following the general excursion set literature, i}n \citet{le2012} and \citet{ccl2013}, Eq.~(\ref{eq:revolution2}) is recast to a simpler form by making several changes of variable: they define $N \equiv \ln(a)$ and $y_j(t)\equiv r_j(t)/a(t) R_j$, where $R_j$ is the initial comoving radius. Derivatives with respect to $N$ are denoted by $y' = \de y / \de N$. We do not repeat in detail here, but only show the result from those references:
\begin{eqnarray}\label{eq:yevolution2}
y''_{j} + \left[2-\frac{3}{2}\Omega_m(N)\right]y'_{j} + \frac{\Omega_m(N)}{2} \, [y_{j}^{-3} (1 + \eta) -1] y_{j} = 0 \, ,
\end{eqnarray}
with initial conditions {specified at $a_{\rm ini}\ll1$ (deep into the matter dominated epoch), as}
\be \label{eq:bcy}
y_j(a_{\rm ini})=1-\delta_{j,\rm ini}/3, \quad y'_j(a_{\rm ini})=-\delta_{j,\rm ini}/3 \, .
\ee
Note that in the above, $\Omega_m(N) \equiv \Omega_me^{-3N}/(\Omega_me^{-3N}+\Omega_\Lambda),$ and 
$\Omega_\Lambda(N) \equiv \Omega_\Lambda/(\Omega_me^{-3N}+\Omega_\Lambda)$. $\delta_{j,{\rm ini}}$ is the initial density perturbation of the spherical patch ($j={\rm in}$) or its environment ($j={\rm env}$).

Thus, the solution is given by considering two second-order differential equations, Eq.~(\ref{eq:yevolution2}) with $j = [{\rm in}, {\rm env}]$ along with the boundary conditions given in Eq.~(\ref{eq:bcy}). {The force ratio $\eta$ in Eq.~(\ref{eq:eta-evo}) at arbitrary time is obtained by interpolating the table (cf.~Fig.~\ref{fig:force-table}) for both halos (over-densities, cf.~Fig.~\ref{fig:force-table}) and voids (underdensities), which is straightforward provided the values of $r/\lambda_{\rm out}$ and $\lambda_{\rm out}/\lambda_{\rm in}$ at that time.}

\section{Void Assignment in the Excursion set: Lagrangian vs.~Eulerian}
\label{sect:method}

In this section we first briefly review some of the key aspects of a recently 
proposed void assignment formalism based on Eulerian arguments (in contrast to the 
traditional Lagrangian-based approach). 
We will then describe how to extend this Eulerian-based void assignment to 
modified gravity theories. For clarity we will choose a single set of
parameters $(\alpha, 2\gamma^2) = (10^{-6},1/3)$ to demonstrate the formalism{, and in a future work we plan to apply the methodology developed in this work to study void abundances in a class of chameleon-like theories}.

\subsection{Void assignment in excursion set formalism}
\label{sec:plsGR}

The excursion set formalism \citep{bcek} 
was initially applied to describe the abundance of halos: the calculation is 
mapped to the computation of the first crossing distribution across some prescribed 
barriers \citep[see, for example, ][for a review]{zentner2007}.
In its simplest form, the first crossing across 
a constant barrier whose amplitude is obtained from the spherical 
collapse model is evaluated for random walks whose heights depend on the linearly 
extrapolated matter power spectrum as well as the smoothing window filter:
analytic solutions are available for sharp-$k$ filter \citep{zh2006,lamsheth09}.
Recent developments improve the model by including scale-dependent \citep{st1999,st2002} 
barriers, diffusive barriers \citep{mr10b,ca11},
random walks with correlated-steps \citep{mr10,pls2012a,ms12}, and
peak constraints \citep{ps2012,psd2013}.

The formation of voids, the biggest under-dense regions in the Universe, 
can also be formulated in the excursion set formalism.
\citet{sv2004} demonstrated that this calculation has to include an additional 
criterion in order to avoid over-counting the number of voids: the so-called 
void-in-cloud effect in which 
voids sitting in over-dense {regions where halos are forming} should be excluded.
Technically, this is realised in the excursion set formalism by imposing 
two barriers for halo {(denoted by $\delta_c$)} and void {(denoted by $\delta_v$)} formation respectively. One then computes 
the first down-crossing probability across the void formation barrier of random walks
that \emph{never} crossed the halo formation barrier at larger smoothing scales.
\citet{sv2004} provided an analytical solution for the case where both halo and void barriers are constant; 
\citet{lsd2009} generalised the solution to barriers of arbitrary shapes.
Some more recent attempts of formulating void abundance include 
Eulerian-void assignment \citep[more details below]{pls2012b}, modifying the mapping from 
Lagrangian volume to abundance \citep{jlh2013}, and introducing diffusive barriers \citep{anp2013}.
In this work we will focus on extending the Eulerian-void assignment in a context in which gravity is modified.

The main purpose of the Eulerian-void assignment is to look for the biggest Eulerian volume that 
satisfies the void criteria: having a density below 20\% of the background.
Conservation of mass  requires the Lagrangian patch to expand 5 times (in volume).
Another essential assumption in the formalism is no shell crossing: while the 
relative separations change, concentric shells (and hence the mass within them) 
preserve their orderings \citep[see, for example,][for discussion]{sheth98}. 
Hence the immediate environment would have significant impact on the formation of voids:
the void-in-cloud described above is a special case in which the surrounding environment collapses
into a vanishing Eulerian size. (Note that it is only a limiting case in which the 
spherical collapse approximation results in vanishing volume. 
In reality halos are virialised objects with physical sizes). 
However, this is only part of the picture: halo-forming is not a necessary condition to 
modify the formation of voids. 
For example, if an under-dense region is embedded in a slightly over-dense environment, 
the comoving volume of this slightly over-dense environment would  
decrease and restrict the expansion of the under-dense patch. 
Whether a void would be formed depends on the \textit{Eulerian} size 
of the surrounding over-density environment. There are three possibilities:
\begin{itemize}
\item A void of the same size as predicted by the Lagrangian treatment {\citep{sv2004}} will be formed;
\item A void will be formed, however its size is smaller than its Lagrangian 
      treatment counterpart;
\item No void will be formed (in other words 
       the limiting case where the void size is zero).
\end{itemize}
As a result, the Eulerian-volume void assignment consists of
a remapping of the size of the void being formed 
and the volume function 
of voids is modified accordingly. For clarity we refer to this Eulerian void 
assignment as the PLS \citep{pls2012b} void assignment in the following.

The evolution of the Eulerian volume of the surrounding environment is a key 
ingredient in the PLS void assignment.  In \citet{pls2012b},
they use the following spherical collapse approximations \citep{b1994,sheth98}:
\begin{eqnarray}
\Delta_{\rm NL} \equiv \frac{M}{\bar{\rho} V_{\rm E}} 
     \approx \left(1 - \frac{\delta_l}{\delta_c}\right)^{-\delta_c}
\end{eqnarray}
to obtain the relationship between the Lagrangian volume (equivalently, 
the total enclosed mass $M$) and the linearly extrapolated density contrast {($B_{V_{\rm E}}=\delta_l$) for a given Eulerian volume $V_{\rm E}$}:
\begin{eqnarray}
B_{V_{\rm E}} (M) = \delta_c \left[1 -  \left(\frac{M}{\bar{\rho} V_{\rm E}}\right)^{-1/\delta_c}\right].
\label{eqn:EulerianB}
\end{eqnarray}
$B_{V_{\rm E}}(M)$ defines a barrier in the $\delta_l - s$ plane for 
Eulerian volume $V_{\rm E}$, 
where $S$ is the variance of the smoothed linear power spectrum over a 
Lagrangian volume $V_{\rm L}$ which satisfies $M = \bar{\rho}V_{\rm L}$,
\begin{eqnarray}
s \equiv \sigma^2(M) = \int \frac{{\rm d}^3k}{(2\pi)^3} P(k)W^2(kR_{\rm L}),
\end{eqnarray}
in which $W(k R_{\rm L})$ is the Fourier transform of the smoothing window function and 
$R_{\rm L}$ is the Lagrangian radius associated with the Lagrangian volume $V_{\rm L}$.

Note that Eq.~\eqref{eqn:EulerianB} defines a monotonically decreasing barrier 
in the $\delta_l - s$ plane  and its limiting value for $\Delta_{\rm NL} \rightarrow \infty$ 
is $\delta_c$. When $\Delta_{\rm NL}=0.2$ and $\delta_c= 1.676$, 
{$B_{V_{\rm E},~\Delta_{\rm NL}=0.2}(M) \approx -2.76$} which is the
constant void formation barrier. \citet{sheth98,lamsheth08a}
discussed the application of the excursion set with barrier defined in
Eq.~\eqref{eqn:EulerianB} to obtain the mass distribution at some fixed
Eulerian volume.

Changing $V_{\rm E}$ in Eq.~\eqref{eqn:EulerianB} results in a series of nested 
curves whose heights are lower for 
increasing $V_{\rm E}$. The limiting case where $V_{\rm E}\rightarrow 0$ is a constant barrier 
$B_{V_{\rm E}} = \delta_c$ and this corresponds to the halo formation barrier.
The PLS algorithm looks for the biggest Eulerian volume which has 
a density contrast of $-0.8$: we are looking for random walks that
cross some Eulerian barriers below the threshold $\delta_v= -2.76$. 
For each random walk, we start from a very large Eulerian volume.
Since the Eulerian barrier associated with a big volume is decreasing
very rapidly, the random walk almost always pierce that barrier above
$\delta_v$. As we gradually decrease the Eulerian volume, the
flattening of the barrier allows the random walk fluctuates before piercing 
the barrier. Ultimately one may find an Eulerian volume whose barrier is
only pierced by the random walk below $\delta_v$ and this Eulerian
volume is the Eulerian size of the void associated with the random
walk. 

\begin{figure}
  \centering\includegraphics[width=0.5\textwidth]{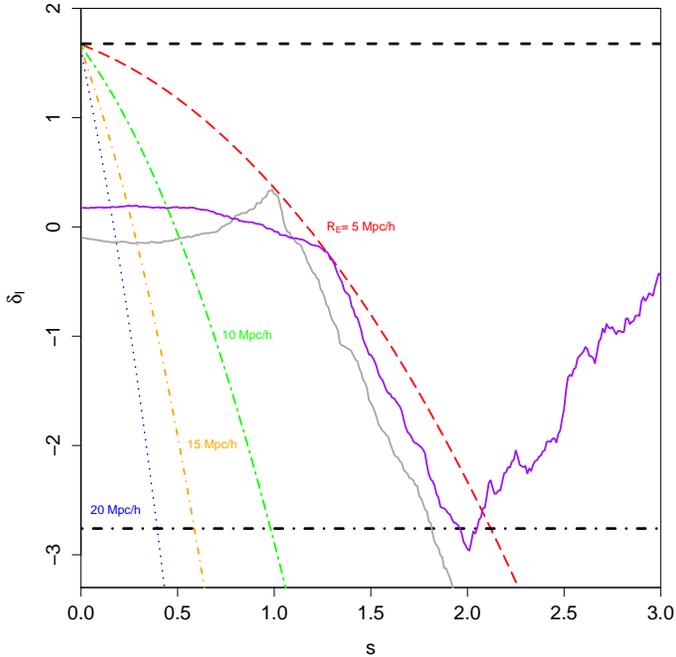}
  \caption{Illustration of Eulerian-void assignment. Barriers for four different
           Eulerian volumes are shown, together with the constant barriers for 
           halo (top thick dashed) and void (bottom thick dot-dashed) formation.
           Two sample random walks shown demonstrate the characteristics 
           of Eulerian vs Lagrangian void assignment (see text for descriptions).
           }
  \label{fig:eulerV}
\end{figure}
Figure~\ref{fig:eulerV} uses two sample random walks to illustrate the characteristic 
of PLS void assignment 
(note: these walks are for demonstration purposes and while
non-vanishing heights at $s=0$ are unphysical, they 
do not affect our discussion). 
Two thick horizontal curves are halo (top) and void (bottom) 
thresholds in GR. 
Four decreasing barriers show Eulerian barriers for different Eulerian
sizes. Both walks represent void forming regions in Lagrangian-void assignment:
they cross the bottom barrier (at approximately $s=1.8$ and $2$ respectively) 
without crossing the upper constant barrier at smaller $s${, and as such they represent voids of similar sizes in the Lagrangian void assignment scheme}.
The result is very different when Eulerian-void assignment is considered:
the grey walk is going to form a void of Eulerian radius of $5$ Mpc$/h$ 
while the purple walk is not a void (in the regions shown).
The red decreasing barrier corresponds to an Eulerian size $R_{\rm E}=5$~Mpc$/h$. 
The grey walk touches it at around $s=1$ but the walk never pierces this barrier above $\delta_v$. To see it is the biggest Eulerian size allowed for the grey random walk, 
imagine we increase the Eulerian size by $\Delta R_{\rm E}$.
The Eulerian barrier associated with $R_{\rm E} + \Delta R_{\rm E}$ lies somewhere between 
the green and red barriers -- the grey random walk will pierce such barrier 
with $\delta_l > \delta_v$, and as a result the density contrast within  
$R_{\rm E}+\Delta R_{\rm E}$ will be higher than 20\% of the background.
This sudden increase in mass corresponds to the 
wall around the void as
discussed in \citet{pls2012b}.

The purple random walk has a different story: although it 
crosses the $\delta_v$ barrier without crossing the $\delta_c$ barrier, the random walk pierces the red Eulerian barrier 
at around $s=2.1$, above the $\delta_v$ barrier. For this particular walk one would 
need to consider smaller Eulerian sizes (and hence the barrier will be shallower) to 
check if a void would be formed (not shown here).

We would like to point out two important characteristics of this PLS 
void assignment as compared to the Lagrangian version:
\begin{enumerate}
\item Random walks identified as void-forming in Lagrangian-void assignment 
      (crossing $\delta_v$ at some $S$ but never crossed $\delta_c$ for $s < S$)
      may not form a void.
      However the reverse is true: PLS assigned voids are always associated with 
      a Lagrangian counterpart.
\item PLS assigned voids cannot be bigger than the corresponding 
      Lagrangian-assigned voids.
\end{enumerate}

\subsection{Void assignment in modified gravity models}
\label{sec:voidES}

The above discussion on Eulerian-void assignment 
applies to models where the gravity is described by general relativity. 
In this section we describe how to extend the various
void assignments to modified gravity models. 
\citet{ccl2013} discussed the extension of Lagrangian-void 
assignment and found that the abundance of voids would constrain
modified gravity models. In particular, the signature of modified gravity is stronger 
at the bigger voids, which also justifies their approximation of neglecting the 
void-in-cloud effect. 
In what follows we review how to apply the Lagrangian void assignment in modified gravity models as well as describe how to apply the PLS void assignment.

In models with modified gravity the dynamics is modified by the presence of the fifth force, 
whose strength can depend on the environmental density. Thus 
all the barriers involved are modified accordingly.
There are multiple scales involved in this case: an environmental scale corresponding to 
halo formation, as well as various Eulerian volumes and their corresponding environments against 
which the void formation criteria is checked.
As discussed in \citet{le2012}, the environment should be big enough to 
encompass the objects being considered, 
but at the same time give a representative environment to that particular object (hence it 
cannot be too large). \citet{lilam2012,lamli2012} proposed using an Eulerian size of 
$R_h = 5$~Mpc$/h$ {(which was roughly the Compton wavelength of the scalar filed in the models considered)} as the environment for halo formation and we will adopt the same here.
For void formation, \citet{ccl2013} suggested a scale 5 times the void radius as the environment, each we also adopt here. 
While Lagrangian environment is used in \citet{ccl2013}, we use Eulerian environment
in the following since it allows us to follow the evolution of the Eulerian volumes closely.
Notice that {previous} studies always assume the evolution of the environment 
follows that of GR since it is always much bigger than the 
Compton wavelength of the scalar field being considered. 
We will also make this assumption in the following discussion.

We apply the PLS void assignment as follows. 
We first generate the following tables 
of different barriers for different environmental densities $\delta_{\rm env}$ {(here $\delta_{\rm env}$ is the linear density contrast of the environment, extrapolated to today using the $\Lambda$CDM linear growth factor; $\delta_{\rm env} > 0$ denotes overdense environments and vice versa)}:
\begin{enumerate}
\item Halo formation barrier (in GR it is a constant barrier at $\delta_c$);
\item Void formation barrier (in GR it is again a constant barrier at $\delta_v$);
\item Eulerian barriers $B_V(S|\delta_{\rm env})$ for 
       each Eulerian volume $V_{\rm E}$.
\end{enumerate}

\begin{figure}
\centering\includegraphics[width=0.49\textwidth]{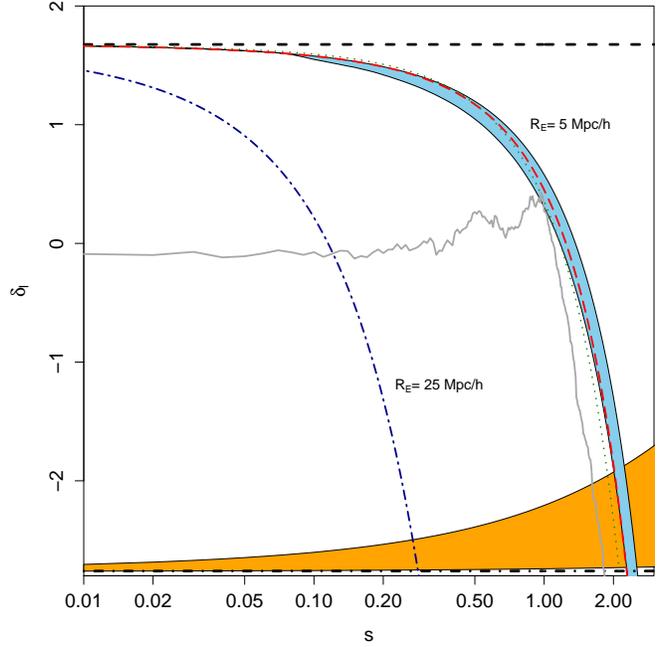}
\caption{Illustration of the PLS void assignment in modified
  gravity models (see text for description).
        The constant barriers at the top and the bottom are the halo and void formation 
         threshold respectively in GR. The orange shaded region is the void formation 
         barriers in modified gravity model for $-2.4 \leq \delta_{\rm env} \leq 1.5$.
         Eulerian barriers for $R_{\rm E}=5$ Mpc/$h$ are shown: 
          the red curve going through the blue shaded region is 
         the corresponding GR barrier while the blue shaded region shows the modified gravity barriers 
         with the same range of $\delta_{\rm env}$ as in the void 
         formation. The green dotted curve shows the spherical collapse 
         approximation [equation~\eqref{eqn:EulerianB}] in GR.
        }
\label{fig:barriersMG}
\end{figure}

Although we assume that the evolution of all the environments would be described by GR,
we use the same numeric solver to obtain the GR Eulerian volume barrier 
for consistency. Figure~\ref{fig:barriersMG} shows results for 
different barriers. For an Eulerian size of $5\ {\rm Mpc}/h$, the blue-shaded
region corresponds to its Eulerian barriers for a wide range of environmental
density contrast ($-2.4 \leq \delta_{\rm env} \leq 1.5$).
The red dashed curve running from the top of the 
shaded region at small $s$ to the 
bottom of the shaded region at big $s$ is the GR Eulerian barrier computed 
by the same numerical solver. 
On the other hand the green dotted curve shows the approximation formula 
Eq.~\eqref{eqn:EulerianB}.

\begin{figure}
\centering\includegraphics[width=0.49\textwidth]{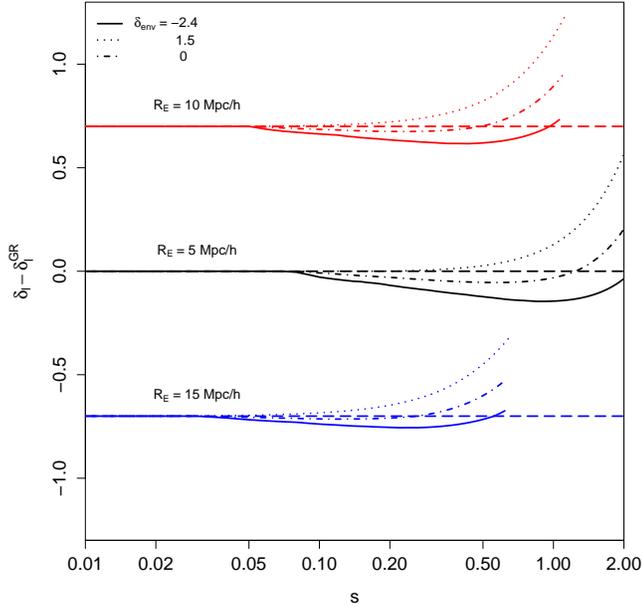}
\caption{Difference between the modified gravity and GR Eulerian barriers  
         for $\reul = 5,~10,~15~{\rm Mpc}/h$  for three different
         environmental overdensity. The results for $\reul = 10$ and 
         $15$~Mpc$/h$ are shifted by $\pm 0.7$. We truncate 
         the comparison for $\delta_l < -3$.
          }
\label{fig:bENV}
\end{figure}

Figure~\ref{fig:bENV} shows the difference between
barriers of three different Eulerian
sizes from the GR counterparts. Three 
environmental density contrasts ($\delta_{\rm env}= -2.4,1.5,0$) are
chosen for comparison, and the result shows the following features:
\begin{itemize}
\item At very large scales (small $s$), there is no difference between the GR and modified gravity barriers. At larger $s$, the two deviate from each other: the deviation starts at smaller $s$ for bigger Eulerian size {since, with $\delta_{\rm env}$ fixed, to have a bigger Eulerian size the initial density $\delta_l$ is in general lower (cf.~Fig.~\ref{fig:barriersMG}), which means the effect of the fifth force is stronger}. 
\item Whether the Eulerian barriers in modified gravity are higher or lower than their GR counterparts depends on the relative density contrast to the environment. If the density inside is higher than that in the environment, the fifth force is attractive (i.e., points towards the centre) and slows down the expansion. To have the same $R_{\rm E}$ today the patch needs to have a lower initial density which helps to speed up the expansion. If the density inside is lower than that in the environment, the fifth force helps the expansion, which has to be compensated by having higher initial densities in the patch. A special case is when $\delta_l=\delta_{\rm env}$, so that the fifth force vanishes identically throughout the evolution: here $\delta_{l}=\delta_l^{\rm GR}$, and is where the curves cross zero in Fig.~\ref{fig:bENV}. As an example, taking the dotted curves in Fig.~\ref{fig:bENV}, which correspond to a very dense environment ($\delta_{\rm env}=1.5$) so that the barrier is higher in modified gravity for most values of $s$; however, when $s$ is extremely small, the initial size of the patch is huge and for it to evolve to a given $R_{\rm E}$ today its initial density has to be higher than 1.5, so that the patch must be over-dense. In this case, the modified gravity barrier becomes lower than in GR. This trend is followed by all curves in Fig.~\ref{fig:bENV}, which only differ by their values of $s$ at which $\delta_l-\delta^{\rm GR}_{l}$ crosses zero.
\end{itemize}

{To better understand the physics, we shall build up the PLS void assignment scheme step by step, following three steps.}

\subsubsection{Void barrier (1-barrier)}

As discussed in \citet{ccl2013}, we fix the environment to be five times 
the size of the resulting voids. The void formation critical density depends
on the environmental density at the corresponding environment. 
The first crossing distribution is therefore
\begin{eqnarray}
f(S) = \int_0^S {\rm d}s\ f(\delta_{\rm env},s;R_{\rm env}) f(S,\delta_v(\delta_{\rm env})| s,\delta_{\rm env}),
\end{eqnarray}
in which $f(\delta_{\rm env},s;R_{\rm env})$ is the first crossing probability of 
the environmental barrier of Eulerian size $R_{\rm env}$ that 
corresponds to a void of size $S$ -- it is five times the void size. 
$f(S,\delta_v(\delta_{\rm env})| s,\delta_{\rm env})$ is the conditional 
first crossing probability of the modified void formation criteria density,
after having first crossed the environment Eulerian barrier at 
$(s,\delta_{\rm env})$. Notice that one does not impose 
the restriction of not crossing the halo formation barrier here.

\subsubsection{Void-in-cloud (2-barrier)}

The void-in-cloud effect is straight-forward to implement: 
the first crossing across
an Eulerian barrier corresponding to a halo formation environment of
$R_h= 5\ {\rm Mpc/h}$ is obtained.
The resulting $\delta_{\rm env}^{\rm halo}$ 
will be used to compute the halo formation barrier.
We then check that the random walks never cross this halo formation barrier 
to enforce the void-in-cloud condition.

\subsubsection{PLS void assignment}

The implementation of the Eulerian-void requires more attention and we will illustrate
the idea with the following example. For simplicity assume that we choose a random walk that 
has not crossed the halo formation barrier. 
At scale $s$ the random walk has a height $\delta_l$: 
we would like to get the Eulerian volume $V_{\rm E}$ at \emph{this point}, and at the moment 
we neglect previous steps of this walk.
In the case of GR, given $M$ and $B$, 
one can find the associated Eulerian volume either by 
looking up the table or by
inverting Eq.~\eqref{eqn:EulerianB}.
In the case of modified gravity models, the exact Eulerian volume $V_{\rm E}$
associated with $(s,\delta_l)$ depends on the environmental density{, while the size (and density) of the environment depends on $V_{\rm E}$, because we have defined void environment to have 5 times the size of the void that forms inside it}. We use the following iterative process to {break this inter-dependency and} obtain $V_{\rm E}$:
\begin{enumerate}
\item Take a guess for the Eulerian size $R_{\rm E}$ (for example, using the corresponding GR value);
\item We now fix the parameters $s$, $\delta_l$, and $R_{\rm E}$ and vary 
      $\delta_{\rm env}$.
\item Set $R_{\rm env} = 5 R_{\rm E}$ and use the 
      GR numerical table to construct the associated environmental barrier
       (recall that we assume GR applies in the environment);
\item The first crossing of the environmental barrier associated with 
      $R_{\rm env}$ gives $\delta_{\rm env}$;
\item We now fix three parameters: $s$, $\delta_l$, and $\delta_{\rm env}$, and vary $R_{\rm E}$;
\item We then search the table of Eulerian barriers in modified gravity models, 
       $B_{V_{\rm E}}(S|\delta_{\rm env})$,
      for an Eulerian volume $V_{\rm E}'$ that satisfies the three parameters. 
      This gives a new estimate of $R_{\rm E}'$;
\item Repeat step {(iii)} until we finally arrive at a consistent set of 
      $\{s,\delta_l,R_{\rm E},\delta_{\rm env}\}$. {This is deemed to be achieved when $R_{\rm E}$ changes by less than 1\%, or less than 0.5 Mpc/$h$ in absolute value, between two consecutive trials, and we call this the convergence of $R_{\rm E}$.}
\end{enumerate}
The resulting environmental density contrast $\delta_{\rm env}$ 
is then used to obtain the associated modified gravity void formation barrier.
This iterative process is applied at each step of the random walk. We then look for the biggest Eulerian volume that the random walk crossed below the corresponding void formation barrier.

Figure~\ref{fig:barriersMG} uses a sample
random walk to illustrate the idea. 
We will take the value $s \approx 1$ where the random walk touches 
the red dashed curve as a demonstration.
At this particular value of $s$, the random walk has a height of around
$\delta_l = 0.3$.
We now need to find the corresponding Eulerian size for \textit{this}
particular value of $(s,\delta_l)$. 
In the case of GR, this Eulerian size would be $5\ {\rm Mpc}/h$. 
Using this initial guess we obtain an environmental size of $R_{\rm env} = 25\
{\rm Mpc}/h$, whose (GR) Eulerian barrier is shown by the blue
dot-dashed curve. The  first crossing of this environment barrier
is approximately $\delta_{\rm env} = -0.1$. We then use this
$\delta_{\rm env}$ value to find a new $\reul'$ whose Eulerian barrier 
will have the value $(s,\delta_l)$. 
In this particular case the Eulerian barrier for
$(R_{\rm E}, \delta_{\rm env})=(5\ {\rm Mpc}/h,-0.1)$ lies somewhere
in the blue shaded region and hence the new $\reul'$ value will be slightly
larger than $5\ {\rm Mpc}/h$. One then repeat the above procedure
until $\reul$ converges.
This iterative process only gives \textit{one} Eulerian size at 
any particular point in the random walk. To find the largest Eulerian size 
that would form a void, one needs to examine the whole random walk {-- which means that the iteration is done at each step of the walk, with a frequency $\Delta s=0.0025$ for our numerical implementation}.

Notice that the implementation of PLS void assignment is computationally very 
intensive in the case of modified gravity models. For this reason we restrict our algorithm 
to search only for voids having size bigger than $5$~Mpc/$h$ but smaller than 
$100$~Mpc$/h$. We believe this range is appropriate  since smaller voids are difficult 
to identify and previous study suggests that the signature of modified gravity lies in 
the bigger voids. On the other hand voids bigger than $100$~Mpc$/h$ are extremely rare 
and this upper limit should not affect our result.

\section{Results}
\label{sect:results}

{Having outlined our methodology,} in this section we present the results of the void 
abundance in our model of modified gravity. In addition to 
the PLS void assignment scheme we described in the previous section,
we also include comparisons with the Lagrangian-void assignment. 
We will discuss void abundance derived from both correlated and 
uncorrelated steps in the random walks. In order to avoid sample variance when computing void abundance, we apply the different algorithms
on the same set of random walks {in both the uncorrelated and the correlated cases, which is realised by using the same random $k$-mode value $\delta(k)$ but different window function $W(kR)$ in}
\begin{eqnarray}
{\delta(x=0;R) = \int\frac{{\rm d}^3{\bf k}}{(2\pi)^3}\delta(k)W(kR),}
\end{eqnarray}
{to generate the walks. In the uncorrected case, $W(kR)$ is a top-hat filter in $k$ space -- because $\delta(k)$ is a Gaussian random number with different $k$ modes independent of each other, and the $k$-space top-hat filter does not introduce correlations between them, the values of $\delta(x=0,R)$ are uncorrelated when changing $R$; in the correlated case the filter is a top-hat in real space. Note that we can apply the three algorithms (1 barrier, 2 barriers and PLS) to correlated and uncorrelated walks, and therefore there are in total 6 ways to assign voids.} {We run a total of 40 million random walks to get our results. For the PLS method there are 1476740 (GR)  and 2123162 (MG) voids respectively.}

\begin{figure}
\centering\includegraphics[width=0.51\textwidth]{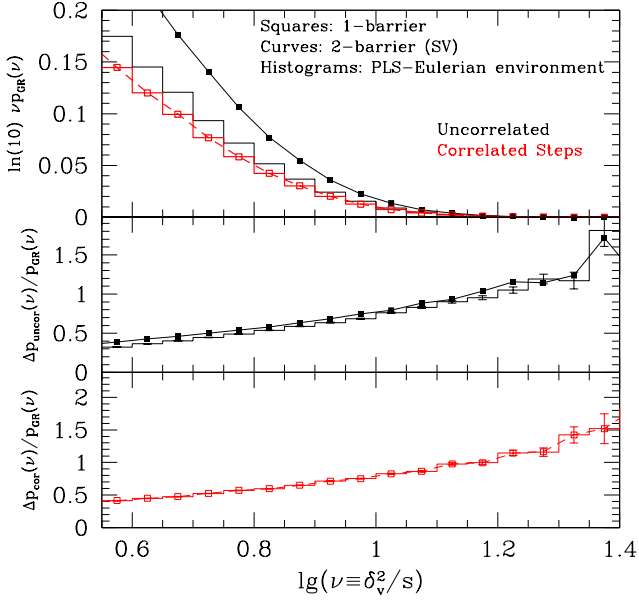}
\caption{The multiplicity function for various void assigning schemes in the 
excursion set approach. The top panel shows various results in GR while 
the middle and the bottom panels show the relative differences in modified gravity models {studied in this paper}.}
\label{fig:pun}
\end{figure}

\begin{figure}
\centering\includegraphics[width=0.51\textwidth]{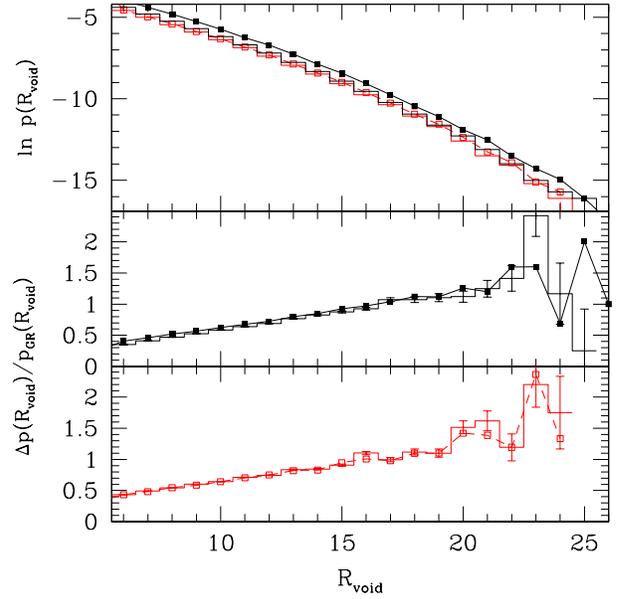}
\caption{The first crossing probability as a function of the Eulerian void 
size for various void assigning schemes in the 
excursion set approach. Same legends are used as in Figure~\ref{fig:pun}.
The top panel shows various results in GR while 
the middle and the bottom panels show modifications in modified gravity models {with uncorrelated and correlated steps void assignment schems}
respectively.}
\label{fig:pr}
\end{figure}

Figure~\ref{fig:pun} shows the first crossing probability 
of various void assignment schemes. The top panel shows 
different predictions for GR while the middle and the bottom panels 
show the differences in modified gravity models.
The black histograms/symbols/curves are the results with uncorrelated steps 
while the red are correlated steps. 
Note that 
in both cases we use the top-hat window function to relate 
the variance $s$ and the smoothing scale in Lagrangian space.

In the case of GR, the three different void assigning schemes have 
very similar results when correlated steps are used {(red symbols)}, showing 
our Monte Carlo simulations are consistent with the results in \citet{pls2012b}.
On the other hand, for the uncorrelated steps,  
the void abundance in the PLS formalism
is significantly lower than the other two.
The results for single barrier (solid black squares)
and 2-barrier (solid curve) are the same, indicating that
at these (relatively large) smoothing scales the void-in-cloud
effect is insignificant {(see below)}. {In other words, in the value range of $\nu$ (which can be mapped to the void radius, cf.~Fig.~\ref{fig:pr}) of interest to us, 
the expansion of voids are indeed restricted by their environments -- the 2-barrier void assignment scheme, 
which only removes voids-in-cloud that eventually collapses, is over-simplistic and not accurate.}

In the middle and the bottom panels we show the modifications 
in void abundance in modified gravity models using uncorrelated steps
and correlated steps in the random walks, respectively. 
We only show the error bars of one of the schemes (in both cases the 
histogram for the PLS method) for clarity.
In the middle panel
the solid squares are the modifications for a single void barrier (hence 
neglecting the void-in-cloud effect) and the solid curve sharing
a very similar result shows the modifications in the 2-barrier case 
(including void-in-cloud effect). Since it is very unlikely 
that these large voids are embedded in even larger patches that are going 
to form halos, the results from these two void assigning methods are 
nearly the same. Notice that it is not straightforward
 to directly compare this panel to 
Fig.~6 in \citet{ccl2013} since we use Eulerian barrier for the 
environment as well as a different linear power spectrum 
generated from {{\sc Camb} \citep{camb}} with the recent Planck cosmology {\citep{Planck}}.
Nonetheless the trend in the abundance modification is very similar in 
both {studies}.
The histogram shows the modifications using the PLS algorithm {with uncorrelated steps}: 
for very big voids the results are similar to the other two void algorithms, 
however there are differences when going to relatively small voids.
The modification in void abundance using the PLS algorithm is slightly less 
than the other algorithms in small voids.

The bottom panel shows the modification of void abundance in modified gravity
models for correlated steps. In this case we employ the top-hat filter
in real space and generate a series of random walks.
The open squares, the dashed curve, and the histogram represent
results from the 1-barrier, 2-barrier, and PLS void algorithm respectively.
All of them show very similar results, suggesting that when using 
correlated steps, one would derive the void abundance using
the much simpler 1-barrier or 2-barrier algorithms. 
We also checked that while there are some quantitative differences, 
the results from uncorrelated steps random walk with 1-barrier or 2-barrier
algorithm agree with the more sophisticated correlated steps with PLS 
algorithm. 

{In summary, while there are noticeable differences in the void abundances from the different 
void assignment schemes, the relative difference between our chameleon model and GR only 
changes slightly. This is somewhat surprising, but it may simply suggest that 
void abundance in chameleon models changes with different voids assignment in a similar manner as in GR.}

Figure~\ref{fig:pr} shows the first crossing distribution plotted 
as a function of the Eulerian sizes of voids being formed. {This figure uses the same set of random walks, but since $s$ and $R$ are not linearly related the walks are organised into different bins.}
We use the same legends as in Figure~\ref{fig:pun}. 
One noticeable difference when comparing the middle and the bottom panels with 
those of Figure~\ref{fig:pun} is stronger modifications at very large voids:
although very rare, the abundance of large voids (larger than 20 ${\rm Mpc}/h$)
is a very sensitive probe of modified gravity models. {Notice that the model parameters used
in this work are mild such that the fifth force helps to enhance the halo mass function by $\lesssim15\%$ across the whole mass range at $z=0$ \citep{le2012}; in contrast, for the void radii studied in this work, we can see that the fifth force can enhance the void abundance by over $50\%$ (and over $100\%$ for $R_{\rm void} > 15$~Mpc$/h$).}

\section{Summary and discussion}
\label{sect:summary}

In this paper we considered the void abundance in {the chameleon-type} modified gravity models --
we have extended previous work in \citet{ccl2013} 
and \citet{pls2012b} to compute the void volume function in modified gravity
models, using the recently developed Eulerian void assignment (PLS) scheme 
within the excursion set approach and to compare its result to other
void assignment schemes.
This Eulerian void method includes the effect of the surrounding on
the growth of void -- it would reduce the size of the void or even 
 {disqualify it} as a void {(if the surrounding collapses)}. A brief description of this PLS void method is 
included in section~\ref{sec:plsGR}.

In order to implement this PLS void assignment method in modified gravity 
models, we solved the spherical collapse equation of motion for various
environmental density contrasts, {to calculate} the critical density contrasts 
for halo and void formation. We performed validation test on the numerical 
solver to make sure the barriers being used in the excursion set approach 
are correct. We then applied these barriers to evaluate the first crossing
probability using Monte Carlo simulations: we used both correlated and 
uncorrelated steps random walks on various void assigning schemes. 
Correlated and uncorrelated steps random walks are associated with 
the window function kernel used to define the Lagrangian patches: the former 
corresponds to top-hat or Gaussian window functions while the latter to 
a sharp-$k$ filter.

We then discussed the implementation of the excursion set approach for 
void volume function in modified gravity models in section~\ref{sec:voidES}.
The 1-barrier and 2-barrier cases are relatively straightforward, 
while we introduce an iterative method in the PLS formalism to 
obtain a consistent description of the parameter set 
$\{s,\delta_l,R_{\rm E},\delta_{\rm env}\}$ -- it is important since 
the evolution of structure depends on the strength of the fifth force, and hence
the environmental density contrast. Notice that these are implemented 
using the Eulerian picture.

The results of our calculation are summarised in 
Figs.~\ref{fig:pun}~and~\ref{fig:pr}.
Our results are consistent with those in \citet{ccl2013}:
the abundances of  voids are very sensitive to modified gravity.
One unexpected finding is the modifications in the void volume function 
due to modified gravity models using the 1-barrier and 2-barrier methods with 
uncorrelated steps match those using the PLS method with correlated steps 
very well. We believe this would be a coincidence -- the fact that the bases of comparison 
(i.e., the GR predictions, see the top panel) are very different supports it\footnote{{Another possibility is that the application of the PLS void assignment method with correlated steps introduces the same differences to the simpler 1-barrier or 2-barrier results, and when taking the relative change this gets cancelled out to a large extent.}}.
On the other hand, the results of the relatively straightforward
1-barrier with correlated steps agree with those of the PLS algorithm, 
for both the GR results as well as the modifications due to modified gravity 
models. Hence when dealing with correlated steps, it is sufficient 
to compute the void abundance using the 1-barrier method -- 
either with Monte Carlo simulations as we did in this work or with
one of the approximations developed for computing first crossing probability.

The enhancement of void abundances due to modified gravity model shows different
characteristic than that of halo abundances: {the chameleon screening is less efficient}
for these under-dense regions and hence the abundances of voids (in particular large ones) are very 
sensitive to modified gravity.
Furthermore, there is detriment in the abundance of low mass halos due 
to mass conservation while the void abundance is always enhanced (down
to the scale we investigate).
By combining the abundances of halos and voids it is possible to break 
the degeneracy between modified gravity models and other cosmological 
parameters or models, such as $\sigma_8$ and massive neutrinos{, a possibility which merits further investigations in the future}.

While void abundance is a sensitive probe of modified gravity models, 
there are several complications needing to be addressed before it can 
be used to put constraints on these models. 
First, real void catalogues are constructed using biased tracers in galaxy surveys{, while our study assumes that voids are found from the dark matter field}.
This may introduce biases -- since modified gravity modifies 
halo abundance (hence galaxies) as well, such biases {may be}
different for GR and modified gravity models.
Second, the predictions from the excursion set approach do not
agree with the void catalogues generated using dark matter particles simulations{, even for the $\Lambda$CDM paradigm}
 \citep[see, for example,][]{jlh2013,anp2013}. 
Different methods are proposed to circumvent this discrepancy but 
it is important to check for consistency  in both GR and modified gravity 
models if any of these methods was used in the comparison analysis.
Third, the increase of void abundance in modified gravity models aggravates
the problem of having a void volume fraction larger than unity. Notice that
the excursion set approach only considers structure formation at 
each position \textit{independently} -- while this may be reasonable for 
collapsing objects (halos), objects (voids) 
having their comoving sizes expanded may have a larger chance to merge with 
each other, invalidating the assumption of the excursion set approach {-- this problem will be worse in the presence of the fifth force, which helps inflate the voids faster. Fourth, given that the fifth force is less efficiently suppressed in voids, the evolution and properties of galaxies there can also be different from in GR, adding more subtlety to the interpretation of observational data and constraints using them.}

Finally, as pointed out in \citet{ccl2013}, since voids sitting in over-dense
environments show stronger signatures of modified gravity models {(at least for the chameleon-type theories)}, one would 
possibly increase the signal-to-noise by using the conditional void volume 
function, or by constructing a weighted void volume function. 
The latter method is now being investigated with both weighted halo mass
function as well as weighted void volume function.

\section*{Acknowledgments}

JC is partially supported by Department of Energy grant de-sc0007901.
YC is supported by the Durham Junior Research Fellowship. 
BL is supported by the Royal Astronomical Society and the Department of Physics of Durham University.
YC and BL acknowledge a grant with the RCUK reference ST/F001166/1.

\label{lastpage}


\begin{thebibliography}{}
\bibitem[\protect\citeauthoryear{Achitouv, Neyrinck \& Paranjape}{2013}]{anp2013} Achitouv I., Neyrinck M., Paranjape A., 2013, arXiv: 1309.3799
\bibitem[\protect\citeauthoryear{Planck Collaboration}{2014}]{Planck} Planck Collaboration, Ade P.~A.~R,~{\it et al.}, 2014, A~\&~A, {\it in press}
\bibitem[\protect\citeauthoryear{Bernardeau}{1994}]{b1994} Bernardeau F., 1994, ApJ, 427, 51
\bibitem[\protect\citeauthoryear{Bond~{et~al}.}{1991}]{bcek} Bond J.~R., Cole S., Efstathiou G., Kaiser N., 1991, ApJ, 379, 440
\bibitem[\protect\citeauthoryear{{Bos}, {van de Weygaert}, {Dolag} \& {Pettorino}}{{Bos} et~al.}{2012}]{Bos2012} {Bos} E.~G.~P.,  {van de Weygaert} R.,  {Dolag} K.,    {Pettorino} V.,  2012, \mnras, 426, 440
\bibitem[\protect\citeauthoryear{{Cai}, {Li}, {Cole}, {Frenk} \& {Neyrinck}}{{Cai} et~al.}{2014a}]{Cai2014a} {Cai} Y.-C.,  {Li} B.,  {Cole} S.,  {Frenk} C.~S.,    {Neyrinck} M.,  2014, \mnras, 439, 2978
\bibitem[\protect\citeauthoryear{{Cai}, {Neyrinck}, {Szapudi}, {Cole} \& {Frenk}}{{Cai} et~al.}{2014b}]{Cai2014b} {Cai} Y.-C.,  {Neyrinck} M.~C.,  {Szapudi} I.,  {Cole} S.,    {Frenk} C.~S., 2014, \apj, 786, 110
\bibitem[\protect\citeauthoryear{Clampitt, Cai \& Li}{2013}]{ccl2013} Clampitt J., Cai Y.-C., Li B., 2013, MNRAS, 431, 749
\bibitem[Clampitt \& Jain(2014)]{cj2014} Clampitt, J., \& Jain, B.\ 2014, arXiv:1404.1834
\bibitem[\protect\citeauthoryear{Clifton~{et~al}.}{2012}]{cliftonetal}Clifton, T., Ferreira, P. G., Padilla, A., Skordis, C., 2012, PhR, 513, 1
\bibitem[\protect\citeauthoryear{Corasaniti \& Achitouv}{2011}]{ca11} Corasaniti P.~S., Achitouv I., 2011, PRD, 023009
\bibitem[\protect\citeauthoryear{Davis {et~al.}}{2012}]{dlss2012} Davis A.~C., Lim E.~A., Sakstein J., Shaw D.~J., 2012, PRD, 85, 123006
\bibitem[\protect\citeauthoryear{{Granett}, {Neyrinck} \& {Szapudi}}{{Granett} et~al.}{2008}]{Granett2008} {Granett} B.~R.,  {Neyrinck} M.~C.,    {Szapudi} I.,  2008, \apjl, 683, L99
\bibitem[Hamaus et al.(2014)]{hws2014} Hamaus, N., Wandelt, B.~D., Sutter, P.~M., Lavaux, G., \& Warren, M.~S.\ 2014, Phys.~Rev.~Lett., 112, 041304
\bibitem[\protect\citeauthoryear{{Hotchkiss}, {Nadathur}, {Gottl{\"o}ber}, {Iliev}, {Knebe}, {Watson} \& {Yepes}}{{Hotchkiss} et~al.}{2014}]{Hotchkiss2014} {Hotchkiss} S.,  {Nadathur} S.,  {Gottl{\"o}ber} S.,  {Iliev} I.~T.,  {Knebe} A.,  {Watson} W.~A., {Yepes} G.,  2014, ArXiv e-prints
\bibitem[Hoyle \& Vogeley(2002)]{hv2002} Hoyle, F., \& Vogeley, M.~S.\ 2002, \apj, 566, 641
\bibitem[Hoyle et al.(2012)]{hvp2012} Hoyle, F., Vogeley, M.~S., \& Pan, D.\ 2012, \mnras, 426, 3041
\bibitem[\protect\citeauthoryear{Hu \& Sawicki}{2007}]{hs2007} Hu W., Sawicki I., 2007, PRD, 76, 064004
\bibitem[\protect\citeauthoryear{{Ili{\'c}}, {Langer} \& {Douspis}}{{Ili{\'c}} et~al.}{2013}]{Ilic2013} {Ili{\'c}} S.,  {Langer} M.,  {Douspis} M.,  2013, Astron.~Astrophys., 556, A51
\bibitem[\protect\citeauthoryear{Jain \& Khoury}{2010}]{jk10} Jain B., Khoury J., 2010, Annals of Physics, 325, 1479
\bibitem[\protect\citeauthoryear{Jennings, Li \& Hu}{2013}]{jlh2013} Jennings E., Li  Y., Hu W., 2013, 434, 2167
\bibitem[\protect\citeauthoryear{Khoury \& Weltman}{2004}]{kw} Khoury J., Weltman A., 2004, PRD, 69, 044026
\bibitem[\protect\citeauthoryear{Lam \& Li}{2012}]{lamli2012}  Lam T. Y., Li B., 2012, MNRAS, 426, 3260
\bibitem[\protect\citeauthoryear{Lam \& Sheth}{2008a}]{lamsheth08a} Lam T. Y., Sheth R. K., 2008a, MNRAS, 386, 407
\bibitem[\protect\citeauthoryear{Lam \& Sheth}{2008b}]{lamsheth08b} Lam T. Y., Sheth R. K., 2008b, MNRAS, 389, 1249
\bibitem[\protect\citeauthoryear{Lam \& Sheth}{2009}]{lamsheth09} Lam T. Y., Sheth R. K., 2009, MNRAS, 398, 2143
\bibitem[\protect\citeauthoryear{Lam, Sheth \& Desjacques}{2009}]{lsd2009} Lam T. Y., Sheth R. K., Desjacques V., 2009, MNRAS, 399, 1482
\bibitem[Lavaux \& Wandelt(2012)]{lw2012} Lavaux, G., \& Wandelt, B.~D.\ 2012, \apj, 754, 109
\bibitem[\protect\citeauthoryear{A.~Lewis}{}]{camb} Lewis A., http://camb.info/
\bibitem[\protect\citeauthoryear{Li}{2011}]{li2011} Li B., 2011, MNRAS, 411, 2615
\bibitem[\protect\citeauthoryear{Li \& Barrow}{2007}]{lb2007} Li B., Barrow J.~D., 2007, PRD, 75, 084010
\bibitem[\protect\citeauthoryear{Li \& Barrow}{2011}]{lb2011} Li B., Barrow J.~D., 2011, PRD, 83, 024007
\bibitem[\protect\citeauthoryear{Li \& Efstathiou}{2012}]{le2012} Li B., Efstathiou G., 2012, MNRAS, 421, 1431
\bibitem[\protect\citeauthoryear{Li, Zhao \& Koyama}{2012}]{lzk2012} Li B., Zhao G.-B., Koyama K., 2012, MNRAS, 421, 3481
\bibitem[\protect\citeauthoryear{Li \& Zhao}{2009}]{lz2009} Li B., Zhao H., 2009, PRD, 80, 044027
\bibitem[\protect\citeauthoryear{Li \& Zhao}{2010}]{lz2010} Li B., Zhao H., 2010, PRD, 81, 104047
\bibitem[\protect\citeauthoryear{Li~{et~al.}}{2011}]{lztk2011} Li B., Zhao G., Teyssier R., Koyama K., 2012, JCAP, 01, 051
\bibitem[\protect\citeauthoryear{Li \& Lam}{2012}]{lilam2012} Li B., Lam T. Y., 2012, MNRAS, 425, 730
\bibitem[\protect\citeauthoryear{Lombriser~{et~al.}}{2012}]{lkzl2012} Lombriser L., Koyama K., Zhao G., Li B., 2011, PRD, 85, 124054
\bibitem[\protect\citeauthoryear{Ma~{et~al.}}{2011}]{maetal11} Ma C., Maggiore M., Riotto A., Zhang J., 2011, MNRAS, 411, 2644
\bibitem[\protect\citeauthoryear{Maggiore \& Riotto}{2010}]{mr10} Maggiore M., Riotto A., 2010, ApJ, 711, 907
\bibitem[\protect\citeauthoryear{Maggiore \& Riotto}{2010}]{mr10b} Maggiore M., Riotto A., 2010, ApJ, 717, 515
\bibitem[Melchior et al.(2013)]{mss2013} Melchior, P., Sutter, P.~M., Sheldon, E.~S., Krause, E., \& Wandelt, B.~D.\ 2013, arXiv:1309.2045
\bibitem[\protect\citeauthoryear{Mo \& White}{1996}]{mw1996} Mo H.~J., White S.~D.~M., 1996, MNRAS, 282, 347
\bibitem[\protect\citeauthoryear{Musso \& Sheth}{2012}]{ms12} Musso M., Sheth R. K., 2012, MNRAS, 423, L102
\bibitem[\protect\citeauthoryear{Oyaizu}{2008}]{oyaizu2008} Oyaizu H., 2008, PRD, 78, 123523
\bibitem[\protect\citeauthoryear{Oyaizu~{et~al.}}{2008}]{olh2008} Oyaizu H., Lima M., Hu W., 2008, PRD, 78, 123524
\bibitem[Pan et al.(2012)]{pvh2012} Pan, D.~C., Vogeley, M.~S., Hoyle, F., Choi, Y.-Y., \& Park, C.\ 2012, \mnras, 421, 926
\bibitem[\protect\citeauthoryear{Paranjape \& Sheth}{2012}]{ps2012} Paranjape A., Sheth R. K., 2012, MNRAS, 426, 2789
\bibitem[\protect\citeauthoryear{Paranjape, Lam \& Sheth}{2012a}]{pls2012a} Paranjape A., Lam T. Y., Sheth R. K., 2012, MNRAS, 420, 1429
\bibitem[\protect\citeauthoryear{Paranjape, Lam \& Sheth}{2012b}]{pls2012b} Paranjape A., Lam T. Y., Sheth R. K., 2012, MNRAS, 420, 1648
\bibitem[\protect\citeauthoryear{Paranjape, Sheth \& Desjacques}{2013}]{psd2013} Paranjape A., Sheth R. K., Desjacques V., 2013, MNRAS, 431, 1503
\bibitem[\protect\citeauthoryear{{Planck Collaboration}, {Ade}, {Aghanim}, {Armitage-Caplan}, {Arnaud}, {Ashdown}, {Atrio-Barandela}, {Aumont}, {Baccigalupi}, {Banday} \& et al.}{{Planck Collaboration}  et~al.}{2013}]{Planck2013} {Planck Collaboration} {Ade} P.~A.~R.,  {Aghanim} N.,  {Armitage-Caplan} C., {Arnaud} M.,  {Ashdown} M.,  {Atrio-Barandela} F.,  {Aumont} J., {Baccigalupi} C.,  {Banday} A.~J., et al. 2013, ArXiv e-prints
\bibitem[\protect\citeauthoryear{Schmidt~{et~al.}}{2009}]{sloh2009} Schmidt F., Lima M., Oyaizu H., Hu W., 2009, PRD, 79, 083518
\bibitem[\protect\citeauthoryear{Sheth}{1998}]{sheth98} Sheth R.~K., 1998, MNRAS, 300, 1057
\bibitem[\protect\citeauthoryear{Sheth \& Tormen}{1999}]{st1999} Sheth R.~K., Tormen G., 1999, MNRAS, 308, 119
\bibitem[\protect\citeauthoryear{Sheth \& Tormen}{2002}]{st2002} Sheth R.~K., Tormen G., 2002, MNRAS, 329, 61
\bibitem[\protect\citeauthoryear{Sheth \& van de Weygaert}{2004}]{sv2004} Sheth R.~K., van de Weygaert R., 2004, MNRAS, 350, 517
\bibitem[Sutter et al.(2012)]{slw2012b} Sutter, P.~M., Lavaux, G., Wandelt, B.~D., \& Weinberg, D.~H.\ 2012, \apj, 761, 187
\bibitem[\protect\citeauthoryear{{Sutter}, {Carlesi}, {Wandelt} \&
  {Knebe}}{{Sutter} et~al.}{2014}]{Sutter2014}
{Sutter} P.~M.,  {Carlesi} E.,  {Wandelt} B.~D.,    {Knebe} A.,  2014, ArXiv
  e-prints
\bibitem[Sutter et al.(2014b)]{spw2014} Sutter, P.~M., Pisani, A., Wandelt, B.~D., \& Weinberg, D.~H.\ 2014, arXiv:1404.5618
\bibitem[\protect\citeauthoryear{Zentner}{2007}]{zentner2007} Zentner A.~R., 2007, IJMPD, 16, 763
\bibitem[\protect\citeauthoryear{Zhang \& Hui}{2006}]{zh2006} Zhang J., Hui L., 206, ApJ, 641, 641
\bibitem[\protect\citeauthoryear{Zhao, Li \& Koyama}{2011}]{zlk2011} Zhao G., Li B., Koyama K., 2011, PRD, 83, 044007
\end{thebibliography}
\end{document}